\documentstyle[12pt]{article}

\def\p{\partial}
\def\g{\gamma}

\def\f2{\frac{1}{2}}

\def\a{\alpha}
\def\b{\beta}
\def\t{\epsilon}

\begin{document}

\title{On the Generalized Hamiltonian Structure of 3D Dynamical Systems}
\author{F.~Haas \& J. Goedert \\
Instituto de F\'{\i}sica, UFRGS\\
Caixa Postal 15051\\
91500-970 Porto Alegre, RS - Brazil}
\date{\strut}
\maketitle

\begin{abstract}
The Poisson structures for 3D systems possessing one constant of motion can
always be constructed from the solution of a linear PDE. When two constants
of the motion are available the problem reduces to a quadrature and the
structure functions include an arbitrary function of them.
\end{abstract}

\section{Introduction}

Three dimensional dynamical systems have deserved much attention both in
view of their intrinsic mathematical relevance and of their wide interest in
domains such as Mechanics \cite{arnold}, optics \cite{dholm}-\cite{puta},
dynamic of interacting populations \cite{lvs}-\cite{epid}, modeling of fluid
turbulence \cite{lorenz}-\cite{kus}, wave interaction models 
\cite{3wav}-\cite{rab}, dynamo theory \cite{rikit}, 
and several other areas of physical,
chemical or biological importance. A more recent related issue concerns the
Poisson structures of these systems. This question has been contemplated
both from the point of view of their existence \cite{perlick} and of their
explicit determination \cite{hojm}-\cite{cong}. As a rule, 3D systems
possess a Poisson structure whenever a sufficient number (e.~g. {\it two})
of independent constants of motion exist \cite{razavy}. Their explicit
construction however is a partly open problem which, as we show in this
letter, can always be solved when two constants of the motion are known.

In a recent paper Gumral and Nutku \cite{gnutku} reduced the problem of the
determination of the structure functions associated to the Poisson structure
of 3D dynamical systems to the solution of a quasi-linear partial
differential equation (eq. (70) of their paper) in three independent
variables. When two independent constants of the motion are known, the
problem therefore reduces to a Ricatti equation. In this letter we show
that in fact the Poisson structures of 3D systems possessing one constant of
the motion can always be obtained from the solutions of a {\it linear}
partial differential equation and that when two independent constants of the
motion are known the problem reduces to a quadrature. In such cases the
resulting structure functions may involve an arbitrary function of the
constants of the motion. When only one constant of the motion is available,
the problem can frequently be handled in the sense that a generalized
Hamiltonian formalism may still be constructed in terms of some particular
solution of the pertaining equations. To illustrate the procedure we shall
consider examples of both types. An interesting example with only one
constant of motion is a five parameter version of the 3D Lotka-Volterra
system for which a Hamiltonian formalism can be constructed. A four
parameter version of the same system was already known to possess a
bi-Hamiltonian structure \cite{nutku}.

\section{Generalized Hamiltonian Structures}

In this section we consider a generic dynamical system in $N$ dimensions 
\begin{equation}
\label{sis-ori}\dot x^\mu =v^\mu ({\bf x},t)\,,\kern 3cm\mu =1,\dots ,N,
\end{equation}
where $v^\mu $ is a sufficiently smooth vector field (in general $v^\mu \in
C^\infty $), ${\bf x}=(x^1,x^2,\dots ,x^N),$ and the over dot denotes
derivative with respect to $t$. In addition we consider a function 
$H({\bf x},t)$ satisfying          
\begin{equation}
\label{hami-pp}\frac{dH}{dt}=\frac{\partial H}{\partial t}\,,
\end{equation}
along any phase trajectory, that is 
\begin{equation}
\label{cond-h}v^\mu \,\partial _\mu H=0\,,
\end{equation}
where $\partial _\mu $ indicates the partial derivative with respect to 
$x^\mu $ and repeated indices represent the Einstein's summation convention.
In specific applications the function $H$ is typically a time-independent
first integral of (\ref{sis-ori}) valid over a region in phase space.
Finally we consider an {\it anti-symmetric\/} matrix ${\cal J}$ which
satisfies the Jacobi identities \cite{olver} 
\begin{equation}
\label{jacobi}J^{\mu [i}\,\partial _\mu J^{jk]}=0\,,
\end{equation}
and therefore provides a generalized definition for the Poisson bracket 
\begin{equation}
\label{p-brac}[F,\,G]\equiv \partial _\mu F\,J^{\mu \nu }\,\partial _\nu G\,,
\end{equation}
of functions $F$ and $G$ in phase space.

\noindent
{\bf Definition:} System (\ref{sis-ori}) is said to be Hamiltonian $iff$
there exists a function $H$ satisfying (\ref{hami-pp}) and an anti-symmetric
matrix ${\cal J}$ satisfying the Jacobi identities such that 
\begin{equation}
\label{hami-form}v^\mu \equiv J^{\mu\nu}\,\partial_\nu H\,. 
\end{equation}
In such case, $H$ is called the Hamiltonian of (\ref{sis-ori}) and ${\cal J}$
the associated Lie tensor or matrix of structure functions.

For given $H({\bf x},t)$ let us label the variables such that $\partial_N H
\neq 0$.

\noindent{\bf Lemma:} if for some anti-symmetric matrix $J^{\mu \nu }$ and 
$H({\bf x},t)$ satisfying (\ref{cond-h}), 
\begin{equation}
\label{eq-lema}
v^s=J^{s\nu }\,\partial _\nu H\,,\kern 1cm\hbox{for}\kern 
1cm s=1\,\,,\dots\,\,,(N-1)
\end{equation}
then 
\begin{equation}
\label{conclus}v^N=J^{N\nu }\,\partial _\nu H\,,\kern 
1cm\hbox{for}\kern 2cm s=N\,.
\end{equation}
The implication of the lemma is clear: if a dynamical system possess a
constant of motion and $N-1$ of its equations are in a Hamiltonian-like form
then its last equation is necessarily of the same form.

Proof: Use (\ref{eq-lema}) to expand (\ref{cond-h}) in terms of $J^{s\nu}$ 
\begin{equation}
\label{proof}\sum_{s\!=\!1}^{N\!-\!1}\partial_s H\,J^{s\nu}\,\partial_\nu H
+ v^N\,\partial_N H = [H,\,H] + \partial_N H\left(v^N-J^{N\nu}\partial_\nu
H\right)=0. 
\end{equation}
The proposition now follows from the last equality, the property of the
Poisson bracket, and the fact that $\partial_N H \neq 0$.

\noindent{\bf Corollary:} When (\ref{eq-lema}) holds, $(N-1)$ components of
the Lie tensor can be represented in terms of the remaining $(N-1)(N-2)/2$
ones: 
\begin{equation}
\label{eq-jNk}J^{N\mu}=- J^{\mu N} = \left(v^\mu -
\sum_{\nu=1}^{N-1}J^{\mu\nu}\, \partial_\nu H\right)/\left(\partial_N
H\right)\,. 
\end{equation}
The proof is constructed by solving (\ref{eq-lema}) for $J^{N\nu}$.

Thus, any function satisfying (\ref{cond-h}) recast system (\ref{sis-ori})
in the ``pre-Hamiltonian'' form (\ref{hami-form}). The algorithm consists of
taking $(N-1)(N-2)/2$ arbitrary functions $J^{\mu\nu} \skip 0.5em
\left(\mu<\nu=1, \dots,(N- 1)\right)$ and filling up the Lie matrix with
them, their anti- symmetric counterpart and the $J^{N\mu}$ given by (\ref
{eq-jNk}). To complete the hamiltonization process we finally demand that 
$J^{\mu\nu}$ obey the Jacobi identities which then become their determining
equations.

In 3D the Jacobi identities are compatible with a generic conformal
rescaling of the structure functions \cite{hojm}-\cite{gnutku}. Such
invariance is, however, partially restricted for $J^{\mu\nu}$ constrained by
relation (\ref{eq-jNk}). In fact, when a constant of the motion exists, the
generic conformal invariance admitted by the Jacobi identities in 3D become
a scale invariance by functions of the constants of motion and the Casimir
only. We now prove this statement:

Suppose that $\bar J^{\mu\nu} \equiv \gamma(x) J^{\mu\nu}$ are structure
functions for some Hamiltonian $\bar H$ when $J^{\mu\nu}$ are the
corresponding structures functions for $H$. The equivalence of the two
representations for the same system implies 
\begin{equation}
\label{equality}J^{\mu\nu}(\gamma\partial_\nu\bar H - \partial_\nu H)=0\,. 
\end{equation}
As a Hamiltonian for a 3D autonomous system, $\bar H$ is necessarily a 
function of the two independent constants of the motion. Locally 
\cite{mscthesis}, these constants may be chosen as the Hamiltonian $H$
and the Casimir of $J^{\mu\nu}$. Therefore, without loss of 
generallity,
\begin{equation}
 \bar H = F(H,\bar C)\,.
\end{equation}
Replacing $\bar H= F(H,\bar C)$ in (\ref{equality})
and considering that the Casimir commutes with all functions of the
dynamical variables yields 
\begin{equation}
\label{gamma}\gamma = 1/\frac{\partial F}{\partial H} = \gamma(H,\bar C)\,. 
\end{equation}

To close the section we remark that in general the Jacobi identities form an
overdetermined system of non linear equations in the unknown $J^{\mu\nu}$.
In fact, when a constant of the motion exists they form a set of 
$N!/\left(3!(N-3)!\right)$ equations in $(N-1)(N-2)/2$ unknowns 
$\{J^{\mu\nu}: \mu\, <\, \nu=1, \dots,(N- 1)\}$. For $N<3$ the Jacobi
identities are automatic and the hamiltonization process is trivial. For $N>3
$ the resulting system of equations is usually overdetermined. For $N=3$
(which is the object of this letter) they become a {\it linear\/} partial
differential equation. When two constants of motion are available, the
problem is reduced to a first order ODE with solutions in terms of
quadratures. The general solution for the structure functions contains an
arbitrary function of the constants of the motion. The prove of this
statement is presented in the following section.

\section{Three dimensional systems}

For $N=3$, equations (\ref{eq-lema}) read 
\begin{eqnarray} 
\label{eq-j3} v^1&=&\hphantom{-}J^{12}\,\p_2 H  + J^{13}\,\p_3 
H \,,\\ v^2&=&-J^{12}\,\p_1 H +J^{23}\,\p_3 H .  
\end{eqnarray} 
We can solve these equations in terms of some function $J$ to be determined
latter 
\begin{eqnarray} 
\label{j123} 
J^{12}&=&J\,,\nonumber\\ 
J^{13}&=&\left(v^1-J\,\p_2 H \right)/(\p_3 H )\,,\\ 
J^{23}&=&\left(v^2+J\,\p_1 H \right)/(\p_3 H )\,.\nonumber 
\end{eqnarray} 
Recall that a different labeling of the variables must be adopted when 
$\partial_3 H \equiv 0$. This implies that equations (\ref{j123}) undergo a
cyclic permutation of the indices ($1 \rightarrow 2 \rightarrow 3
\rightarrow1)$ that exchanges $\partial_3$ with either $\partial_2$ or 
$\partial_1$. This permutation must be applied simultaneously to equations 
(\ref{eq-a}-\ref{eq-b}) bellow.

Equations (\ref{j123}) express all the components of the Lie tensor in 3D in
terms of the symbol $J$ determined by the Jacobi identity. To see this we
substitute (\ref{j123}) into (\ref{jacobi}) and multiply the result by 
$\partial_3 H $. The cross derivative and the quadratic terms cancel out,
and, after using (\ref{cond-h}) and some simple algebra, we obtain 
\begin{equation}
\label{eq-j}v^\mu \partial_\mu J=A\,J +B, 
\end{equation}
where 
\begin{equation}
\label{eq-a}A=\partial_\mu v^\mu - \frac{(\partial_3v^\mu)(\partial_\mu H)}
{\partial_3 H } 
\end{equation}
and 
\begin{equation}
\label{eq-b}B=\frac{v^1\,\partial_3v^2-v^2\,\partial_3v^1}{\partial_3 H }. 
\end{equation}
In this equations the indices $1$, $2$ and $3$ are linked to the indices in
equation (\ref{j123}) and, as already mentioned, should undergo a cyclic
permutation whenever $\partial_3H=0$.

Equation (\ref{eq-j}), which is the Jacobi identity in terms of $J$, is the
key equation in the solution of the hamiltonization problem of 3D systems.
As mentioned before, this last condition is a linear first order PDE. When
two constants of the motion are known, one of the characteristic equations
for (\ref{eq-j}) becomes a linear first order ODE (two out of the three
dynamical variables are solved in terms of a third one and the constants of
motion). In this case $J$ can be solved by quadratures and the structures
functions may always involve an arbitrary function of the two constants of
the motion. 

It is important to remark here that the scale invariance referred to before
is preserved by equations (\ref{eq-j}-\ref{eq-b}): if $H$ is a Hamiltonian
with associated Lie tensor (\ref{j123}) (expressed in terms of $J$) then 
$F(H,C)$, $F$ arbitrary, is another valid Hamiltonian but with associated Lie
tensor $\bar J \equiv J/(\partial F/\partial H)$. In fact, it can be easily
checked that if $J$ satisfies equation (\ref{eq-j}) for $A$ and $B$
calculated from $H$, then $\bar J$ satisfies the same equation with $A$ and 
$B$ calculated from $F(H,C)$. The multiplicity of solutions for the
hamiltonization process is still greater if we consider that equation (\ref
{eq-j}), being a first order linear PDE, in general admits an infinity of
solutions. Sometimes, as we shall show in section 4, solutions in terms 
of arbitrary functions can also be found.  

As final remark of practical interest we point that any particular solution
to (\ref{eq-j}) provides a non trivial Poisson structure for the system
under consideration. Interesting particular cases are obtained when the
system has $B=0$, in which case $J = 0$ is the simplest possible solution or
when $B({\bf x},t) = - f(H,C)\,A({\bf x},t)$ in which case $J=f(H,C)$ is an
equally simple particular solution.

\section{Sample applications}

To illustrate some of the various possibilities offered by the procedure
proposed above we consider now three sample problems in detail. Several
other systems can be equally treated by repeating a similar sequence of
steps. A more exhaustive and detailed list of examples is being prepared and
will soon be submitted for publication elsewhere.

\subsection{The ice skate problem}

As a first example we consider the ice skate problem studied by Lucey \cite
{lucey}. In the original treatment one of the equations was $\dot x^2 = 0$
which can be solved for $x^2 = - a= \hbox{constant}$. This transforms the
4D system into a 3D one. We next relabel the variables using the
replacements $x^4 \rightarrow x^3 \rightarrow x^2$. In this notation the ice
skate system reads 
\begin{equation}
\label{hoj-ex}\dot x^1 = -a\,, \qquad \dot x^2 = x^3\,, \qquad \dot x^3 = a
x^3 \tan{x^1}. 
\end{equation}
We now verify that $H_1 = x^3 sec(x^1)$ is a constant of the motion and
therefore a candidate as the Hamiltonian for the system. Use of this
Hamiltonian as a source in (\ref{eq-a}-\ref{eq-b}) gives $A = 0$ and 
$B=-a\cos x^1$. We next verify that one of the characteristic equations of 
(\ref{eq-j}), 
\begin{equation}
\frac{dJ}{dx^1} = \cos x^1, 
\end{equation}
is separate from the others and can be integrated in the form 
\begin{equation}
\label{ice-j} J =\sin x^1 + F(H_1,H_2)\,, 
\end{equation}
where $F$ is an arbitrary function of constants of the motion. The following
structures functions are now obtained by substitution of (\ref{ice-j}) in 
(\ref{j123}) 
\begin{eqnarray}
J^{12}&=& -J^{21}=\hphantom{ - }\sin x^1 + F(H_1,H_2)\,, \nonumber\\[3mm]
J^{13} &=&-J^{31}= - a\cos x^1\,, \\[3mm]
J^{23} &=&-J^{32}=\hphantom{-}x^3\left(
\sec x^1 +F(H_1,H_2)\tan{x^1}\right)\,.\nonumber
\end{eqnarray}
To find a Casimir of the algebra it is necessary to specify the function 
$F$. In particular for $F = 0\,,$ $C_1=a x^2 +x^3\tan{x^1}\,,$ is a
Casimir of the system.

We now verify that the ice skate problem is in fact completely integrable
since $C_1$ is another constant of motion, functionally independent of 
$H_1$. We may therefore use $H_2=C_1$ as an alternative Hamiltonian in 
which case 
$A=B= -a \cot{x^1}$ and 
\begin{equation}
\label{sec-j}\bar J = - 1 +\bar F(H_1,H_2)\sin{x^1} \,, 
\end{equation}
for any arbitrary function $\bar F$ of the constants of the motion. This
generates the alternative Poisson structure 
\begin{eqnarray}
\bar J^{12}&=& -\bar J^{21}= - 1 + \bar F(H_1,H_2)\sin{x^1} \,, \nonumber\\[3mm]
\bar J^{13} &=&-\bar J^{31}= - a\, \bar F(H_1,H_2)\cos x^1\,, \\[3mm]
\bar J^{23} &=&-\bar J^{32}=\hphantom{-}x^3\left(
- \tan x^1 +\bar F(H_1,H_2)\sec{x^1}\right)\,.\nonumber
\end{eqnarray}
For $\bar F = 0$ the new algebra has Casimir $C_2=H_1=x^3\sec x^1$. It
is worthwhile remarking that the algebras corresponding to $J$ and $\bar J$
are independent and not connected by conformal transformations. The
rescaling invariance by functions of the Hamiltonian and the Casimir can
still be applied to generate families of equivalent Poisson structures.

We have presented a complete and novel solution to the hamiltonization of
the ice skate system. This problem illustrates clearly the basic features of
the routine proposed. In the next subsection we analyze another sample
problem and show that a well known system may still exhibits some novel and
perhaps unexpected features.

\subsection{The Euler top}

We now revisit the Euler top problem and show that also this system
possesses families of alternating Hamiltonian and Casimir with the
associated Poisson structure involving arbitrary functions of the 
two.

The classical Euler top system \cite{arnold} can be written as 
\begin{eqnarray}
\label{euler-t}
\dot x^1 = (I_2-I_3)x^2x^3/(I_2I_3) \,, \nonumber\\
\dot x^2 = (I_3-I_1)x^3x^1/(I_3I_1) \,,\\
\dot x^3 = (I_1-I_2)x^1x^2/(I_1I_2) \,, \nonumber
\end{eqnarray}
and admit the (kinetic) energy 
\begin{equation}
\label{kin-en}H=\frac 12\left( \frac{(x^1)^2}{I_1}+\frac{(x^2)^2}{I_2}+
\frac{(x^3)^2}{I_3}\right) \,,
\end{equation}
and the angular momentum 
\begin{equation}
\label{ang-m}L=(x^1)^2+(x^2)^2+(x^3)^2\,,
\end{equation}
as independent constants of the motion. Taking the energy as the Hamiltonian
we find $B = 0$ and 
\begin{equation}
\label{eq-aet}A=(I_1-I_2)x^1x^2/(I_1I_2x^3)\equiv \dot x^3/x^3\,.
\end{equation}
For these values of $A$ and $B$ the solutions $J$ to the basic equation (\ref{eq-j})
are easily found and the corresponding structure functions become 
\begin{eqnarray}
J^{12}&=& -J^{21}= - x^3\left(1 + F(H,L)\right)\,, \nonumber\\[3mm]
J^{13} &=&-J^{31}= \hphantom{-}x^2\left(1 +I_3 F(H,L)/I_2\right) \,, \\[3mm]
J^{23} &=&-J^{32}=-x^1\left(1 +I_3 F(H,L)/I_1\right)\,,\nonumber
\end{eqnarray}
where again $F$ is an arbitrary function of its arguments. When $F = 0$
the usual results are recovered with the angular momentum $L$ as the Casimir.

Alternatively, starting with the angular momentum $L$ as the Hamiltonian,
leads to the same forms for $A$ and $B$ but the new structure functions
become (where $\bar F$ is also an arbitrary function of its arguments) 
\begin{eqnarray}
\bar J^{12}&=& -\bar J^{21}= x^3\left(1/I_3 + \bar F(H,L)\right)\,,
\nonumber\\[3mm]
\bar J^{13} &=&-\bar J^{31}= - x^2\left(1/I_2 + \bar F(H,L)\right)\,, \\[3mm]
\bar J^{23} &=&-\bar J^{32}= x^1\left(1/I_1 + \bar F(H,L)\right)\,.\nonumber
\end{eqnarray}
For $\bar F = 0$ the kinetic energy $H$ is the Casimir of the new
algebra.

The analysis presented here shows that the Euler top system admits a
continuous family of Poisson structures in which the kinetic energy and the
angular momentum play alternatively the roles of Hamiltonian and Casimir.
Each pair of Hamiltonian and Casimir have associated Lie algebra that are
independent, that is, that are not connected by conformal rescaling.

\subsection{The 3D Lotka-Volterra system}

The 3D Lotka-Volterra system and some of its special subsystems play an
important role in modeling many physical, chemical and biological processes.
Their associated vector field is defined (in their most general form with
Verhulst terms $b_{ii}\neq 0$) by 
\begin{equation}
\label{lvs-comp}v^k \equiv x^k(a_k + b_{k\mu}x^\mu) \hskip 3cm k=1, \dots
N\,. 
\end{equation}
In equation (\ref{lvs-comp}) and throughout the rest of this letter sum over
the Latin index $k$ is not implied.

Cair\'o and Feix~\cite{lvs} found several invariants or first integrals
for N-dimensional Lotka-Volterra systems. In particular for $N=3$ and 
$\det(b_{ij})=0$, one such invariant is 
\begin{equation}
\label{lvs-inv}{\cal H} = H(x^1,x^2,x^3)\,e^{-s\,t} \equiv (x^1)^{\alpha}
(x^2)^{\beta}(x^3)^{\gamma} e^{-s\,t} 
\end{equation}
where $\alpha$, $\beta$, $\gamma$ and $s$ are given by 
\begin{eqnarray*} 
\label{dan34} 
\a &=& b_{22} b_{31} - b_{21}b_{32}\,,\\ 
\b &=& b_{11} b_{32} - b_{12}b_{31}\,,\\ 
\g &=& b_{12} b_{21} - b_{11}b_{22}\,,\\ 
s &=& a_1 \a + a_2 \b + a_3 \g\,.  
\end{eqnarray*} 
Under the constraint $s=0$, that is, for 
\begin{equation}
\label{cond-s}a_1( b_{22} b_{31} - b_{21}b_{32}) + a_2( b_{11} b_{32} -
b_{12}b_{31}) + a_3(b_{12} b_{21} - b_{11}b_{22})=0\,, 
\end{equation}
the function $H$ becomes a time-independent first integral. Bellow we
present the derivation of a Poisson structure for this system under the less
restrictive conditions that we can find. This will ultimately imply that
among the initially free coefficients $a_k$ and $b_{ij}$ only five remain
arbitrary. Unfortunately no other constant of the motion is known for the
same number of free parameters and complete integrability cannot be sought. It
is however interesting to verify that under appropriate conditions one can
construct a Poisson structure without being actually able to integrate the
equations completely.

In order to tackle the basic equation (\ref{eq-j}) we first calculate $A$
and $B$ in terms of the vector field $v^\mu $ and of the constant of the
motion $H$. For this, equations (\ref{eq-a}-\ref{eq-b}) and (\ref{lvs-inv})
yield 
\begin{eqnarray} 
\label{dan17}   
A  &=& a_1+a_2+b_{11}x^1+b_{22}x^2+b_{33}x^3 + (b_{1\mu}  +  b_{2\mu})x^\mu 
\,,\\[5mm] 
B  &=&  \frac{U}{\g H}[(a_1b_{23}-a_2b_{13})+( b_{23}b_{1\mu} - 
b_{13}b_{2\mu})x^\mu]\,,  
\end{eqnarray}  
where we introduced the symbol $U = x^1x^2x^3$ to simplify the notation.
To proceed we write the characteristic equation associated to (\ref{eq-j}) 
\begin{equation}
\label{char}\frac{d\,x^k}{x^k(a_k+b_{k\mu }x^\mu )}=\frac{dJ}{A\,J+B}\,,
\end{equation}
which after some element but tedious algebra implies 
%
\begin{equation}
\label{dan19}\frac{dU}{U\,[A+a_3+(b_{31}-b_{11})x^1+(b_{32}-
b_{22})x^2]}=\frac{dJ}{AJ+B}\,.
\end{equation}
We now multiply the numerator and the denominator of the first term in
equation (\ref{dan19}) by $\epsilon /{\gamma H}$ where $\epsilon $ is an
arbitrary constant to be determined later, and re-write the characteristic
equations in the form 
\begin{equation}
\label{dan20}\frac{d\,x^k}{x^k(a_k+b_{k\mu }x^\mu )}=\frac{d\left[
J-\epsilon U/(\gamma H)\right] }{A\left[ J-\epsilon U/(\gamma H)\right]
-B^{\prime }\left( U/(\gamma H)\right) }\,,
\end{equation}
with $B^{\prime }$ defined by 
\begin{eqnarray} 
\label{bprima}
 B^{\prime} &=&  a_1b_{23}-a_2b_{13}-\t a_3+ \left[\t(b_{11}\!-\!b_{31})
+ b_{23}b_{11}\!-\!b_{13}b_{21}\right]x^1\nonumber\\
&+&\left[\t(b_{22}\!-\!b_{32})\!+\!b_{23}b_{12}\!-\!b_{13}b_{22}\right]x^2\,. 
\end{eqnarray}  
Equation (\ref{dan20}) is difficult to treat in its general form. One
particular solution however can be readily found by imposing the additional
condition $B^{\prime } = 0$. As can be easily checked, under this
condition 
\begin{equation}
\label{lvs-j}J=\frac{\epsilon U}{\gamma H}=\frac \epsilon \gamma
(x^1)^{1-\alpha }(x^2)^{1-\beta }(x^3)^{1-\gamma }.
\end{equation}
satisfies equation (\ref{dan20}) and therefore the fundamental equation (\ref
{eq-j}). The condition on $B^{\prime }$ produces the value of the arbitrary
constant $\epsilon $ and (for arbitrary $x^1$ and $x^2$) two additional
constraints on the coefficients of the system. This represents a total of 
{\it four} constraints ($\det (b_{ij})=0$, $s=0$ and two out of conditions 
(\ref{cond-I}-\ref{cond-III}) below) on the {\it twelve} initially free
parameters. This corresponds to a total of eight free parameters which in
fact reduces to {\it five} since three parameters can always be
eliminated by rescaling.

Condition (\ref{bprima}) for arbitrary $x^1$ and $x^2$, implies three
equations one of which can be solved for $\epsilon$ when either ($i$) 
$a_3\neq 0$ or ($ii$) $b_{31} \neq b_{11}$ or ($iii$) $b_{32} \neq b_{22}$.
These conditions imply the following relations 
\begin{eqnarray}
\label{cond-I}
(a_1b_{23}-a_2b_{13})(b_{31}-b_{11})&=& a_3(b_{23}b_{11}-b_{13}b_{21})\,,\\
\label{cond-II}
(a_1b_{23}-a_2b_{13})(b_{32}-b_{22}) &=& a_3(b_{23}b_{12}-b_{13}b_{22})\,,\\
\label{cond-III}
(b_{23}b_{11}-b_{13}b_{21})(b_{32}-b_{22})&=&(b_{23}b_{12}-
b_{13}b_{22})(b_{31}-b_{11})\,, 
\end{eqnarray}
which we interpret as follow:

\noindent i) when $a_3\neq 0$ apply (\ref{cond-I}-\ref{cond-II}) and use 
\begin{equation}
\epsilon=\frac{1}{a_3}(a_1b_{23}-a_2b_{13})\,; 
\end{equation}
and/or\\ ii) when $b_{31} \neq b_{11}$ apply (\ref{cond-I},\ref{cond-III})
and use 
\begin{equation}
\epsilon=(b_{23}b_{11}-b_{13}b_{21}))/(b_{31}-b_{11})\,; 
\end{equation}
and finally\\ iii) when $b_{32} \neq b_{22}$ apply (\ref{cond-II}-\ref
{cond-III}) and calculate $\epsilon$ from 
\begin{equation}
\epsilon=(b_{23}b_{12}-b_{13}b_{22})/(b_{32}-b_{22})\,. 
\end{equation}

To complete the calculation insert $\epsilon$ in (\ref{lvs-j}) and
substitute $J$ into (\ref{j123}). This provides a Poisson structure for
the 3D Lotka-Volterra system, namely: 
\begin{eqnarray} 
\label{j123-f} 
J^{12} &=& - J^{21} = \frac{\t}    {\g H} x^1 x^2 x^3\,,\\  
J^{13} &=& - J^{31} = \frac{x^1x^3}{\g H}\left(b_{1\mu}x^\mu-
\frac{\t\b}{\g}x^3\right)\,,\\ 
\label{j123-ff}
J^{23} &=& - J^{32} = \frac{x^2x^3}{\g 
H}\left(b_{2\mu}x^\mu+\frac{\t\a}{\g}x^3\right)\,.  
\end{eqnarray} 

In Nutku's analysis \cite{nutku} of the Lotka-Volterra system, one of the
first integrals was the logarithm of the function used above and no Verhulst
(diagonal) terms were included. Like in the present treatment the
determinant of the coefficient of the quadratic terms was set to zero and an
additional constraint was imposed similarly to condition (\ref{cond-s}).
This more constrained system (four free parameters) has a known second
constant of motion (or a constant of motion and a Casimir) 
and consequently admits a bi-Hamiltonian structure. The
component $J^{12}$ of the Lie tensor in Nutku's analysis satisfies condition
(\ref{eq-j}) but is a different particular solution.

\section{Conclusions}

We have presented a procedure for constructing the Poisson structure of
dynamical systems possessing a constant of motion. In three dimensions the
problem is reduced to the solution of a linear PDE. When two
time-independent first integrals of the system are known the procedure can
be applied twice and yields bi-Hamiltonian structures involving 
{\it arbitrary} functions of the constants of motion
(or a constant of the motion and a Casimir). The technique was applied
to a few sample systems to show how it operates. In particular we showed
that the ice skate system analyzed by Lucey \cite{lucey} is completely
integrable, that the Euler top systems admits an infinite family of
bi-Hamiltonian structures and that a five free parameter version of the 3D
Lotka-Volterra system can be cast in a generalized Hamiltonian form. Other
3D systems for which one or more time-independent constants of motion are
known can be equally treated. This is being done for several rescaled
version of such systems found in the literature. The results are still in
preparation (some preliminary results can be found in ref. \cite{cong}) and
will be soon submitted for publication elsewhere.

\bigskip\bigskip 
\leftline{\bf Acknowledgement} \smallskip 
This work was supported by the Brazilian Conselho Nacional de
Desenvolvimento Cient\'{\i}fico e Tecnol\'{o}gico (CNPq). JG thanks M. R.
Feix and L. Cair\'o for fruitful discussions and critical reading of the
manuscript and the staff at PMMS/CNRS (where part of this work was done) for
their kind hospitality.

\end{document}